\documentclass[magazine,comsoc]{IEEEtran}
\IEEEoverridecommandlockouts
\usepackage{cite}
\usepackage{wrapfig}
\usepackage{amsmath,amssymb,amsfonts}
\usepackage{algorithmic}
\usepackage{graphicx}
\usepackage{enumitem}
\usepackage{textcomp}
\usepackage{xcolor}
\usepackage{array}
\usepackage{ulem}
\newcolumntype{P}[1]{>{\centering\arraybackslash}p{#1}}
\def\BibTeX{{\rm B\kern-.05em{\sc i\kern-.025em b}\kern-.08em
    T\kern-.1667em\lower.7ex\hbox{E}\kern-.125emX}}

\begin{document}

\title{Energy-Efficient Intra-Domain Network Slicing for Multi-Layer Orchestration in Intelligent-Driven Distributed 6G Networks: Learning Generic Assignment Skills with Unsupervised Reinforcement Learning}

\author{Navideh Ghafouri , John S. Vardakas, Kostas Ramantas, and Christos Verikoukis% <-this % stops a space
\thanks{Navideh Ghafouri is with Iquadrat Informatica S. L., Barcelona, Spain.}% <-this % stops a spa
\thanks{John S. Vardakas is with Iquadrat Informatica S. L., Barcelona, Spain, and Dept. of Informatics, UOWM, Greece.}% <-this % stops a space
\thanks {Kostas Ramantas is with Iquadrat Informatica S. L., Barcelona, Spain.}
\thanks{Christos Verikoukis is with the CEID, University of Patras, Greece.}
%\thanks {This work is funded by the H2020 research program under the MARSAL project (GA \#101017171) }
}

\maketitle

\begin{abstract}

Since the 6\textsuperscript{th} Generation (6G) of wireless networks is expected to provide a new level of network services and meet the emerging expectations of the future, it will be a complex and intricate networking system. 6G´s sophistication and robustness will be accompanied by complexities, which will require novel strategies to tackle them. This research work focuses on decentralized and multi-level system models for 6G networks and proposes an energy efficient automation strategy for edge domain management and Network Slicing (NS) with the main objective of reducing the networks´ complexity by leveraging scalability, efficiency, and generalization. Accordingly, we propose a pre-train phase to discover useful assignment skills in network edge domains by utilizing unsupervised Reinforcement Learning (unsupervised RL). The suggested technique does not depend on the domain specifications and thus is applicable to all the edge domains. Our proposed approach not only enables scalability and decentralization, but it also delivers efficiency by assisting domain controllers to provide various service types. We implemented the pre-training phase, and monitored that the discovered assignment skills span the entire interval of possible resource assignment portions for every service type. 

\end{abstract}

\section{Introduction}
Throughout the generations of wireless networks, the 5\textsuperscript{th} Generation (5G) appeared distinct by introducing new definitions such as user-tailored service provision. NS has come to be viewed as an important yet complicated procedure in network management and orchestration. This is the result of introducing more layers in the system model of the current and future wireless networks. Since service provision and NS are going to stay and expand in 6G, the dynamicity and robustness of 6G will introduce more complexity to network management and orchestration. However, one way to tackle the complexity of the 6G networks is by considering modular and multi-level system models \cite{0}. While decentralization, scalability, and modularity arise to be a necessity for 6G, NS will become more challenging and delicate. 

Softwarization and programmability are almost recently introduced concepts that enabled smoother and more dynamic network management by transforming traditional hardware-based systems into flexible programmable networks. On the other hand, 6G networks are envisioned to be vigorous and dynamic, eliminating the possibility of managing them with static techniques. 

Moreover, the real-time performance of 6G entities and their orchestration introduce real-life kind challenges that confirm the engagement of Artificial Intelligence (AI) in future wireless networks. Reinforcement Learning (RL) has been employed as a viable network management and orchestration strategy. RL-based techniques appear in various approaches for decentralized problem models, including Multi-Agent RL (MARL) and Multi-Objective RL (MORL). While in MARL multiple agents can cooperate or oppose each other to reach a goal in MORL one agent is trained for multiple objectives to consider as a set of goals. In addition to RL, Federate Learning  (FL) is another Machine Learning (ML)-based technique that enables training models across multiple decentralized devices or servers holding local data, without the need to centralize the data in one location.
Network orchestration, management, and NS by utilizing AI-based techniques with the help of network softwarization have been the subject of many research works lately \cite{1,2,11,12,13}.

Nevertheless, a sophisticated network orchestration model in real scenarios can be limited by its complexity. Therefore, as 6G becomes more tangible, practicality and scalability can be considered as the new concerns to focus on when developing novel ideas for 6G networks. In addition to complexity, how the network impacts the environment is a critical subject to take into account. Accordingly, deploying ML-based management approaches that require less computational resources and, thus, are more energy efficient is necessary. 

In this research work, we follow the purpose of developing management and orchestration techniques that are efficient and scalable. We propose a novel framework for the reduction of complexity and increase of efficiency for the automated edge network domains. This framework consists of an offline pre-training phase that discovers useful assignment skills for the domain controller. In this phase, an unsupervised RL agent discovers useful sequences of assignments for various resource types. The domain controller, which provides the service types, uses the discovered skills in the next phase and manages NS inside the edge domains. The following aspects motivated us to propose learning assignment skills in an unsupervised RL-based pre-training phase for 6G orchestration:
\begin{itemize}
    \item The network system model is modular and consists of network domains with the same nature but different details. In order to decrease the complexity and consequently improve scalability and efficiency, we investigate automation approaches that are not limited to one network domain (environment). 
    \item The management system in each domain is faced with multiple tasks that are assignments of various service types. In other words, the process of assigning a network slice is a different procedure for each service type for the controller. We are interested in controllers that are capable of performing multiple tasks yet do not require complicated, costly techniques. Therefore, we follow the purpose of providing multiple service types with only one undemanding controller.
    \item The network domains can be observed as complex environments for intelligent controllers. By discovering useful skills in an offline pre-training phase, not only do we reduce the controller´s action domain noticeably, but we also wrap detailed information and present the useful information in an optimal way. By using useful assignment skills for service-providing tasks instead of all possible options, we filter the options with a better chance of effectiveness, which increases the controller´s success rate significantly. 
    \item The management and orchestration of the 6G system model needs to be modular and scalable. Hierarchical orchestration levels are very likely to exist in these systems. Developing practical assignment skills through the offline phase proves adaptable and beneficial for overall system orchestration and management.
    \item The proposed approach is scalable and efficient. The generic unsupervised RL-based skill discovery approach can be used to automate any number of network domains with the minimum cost of deploying the same controller with the offline discovered skills for each domain without any scalability issues.
\end{itemize}

In the rest of this article, we present some major transitions in networking when entering the 6G era, and we present our considered system model for edge domain orchestration in 6G networks. Later, We have a short review of the enabling technologies and our methodology. Next, we introduce our generic unsupervised RL-based offline skill discovery approach. Finally, the potential applications and prospective future developments of our proposed approach are presented.

\section{from network softvarization and conventional communication-oriented management to fully intelligent networks with task-oriented approaches}
While in the 5G era, we have witnessed the emergence of revolutionary concepts such as softwarization and NS, AI will help the 6G networks to fully take advantage of these concepts. In addition to softwarization and NS, Software-Defined Networking (SDN), Network Function Virtualization (NFV), and NS have moved modern communications toward software-based virtual networks, targeting to meet the 6G era requirements. Adding intelligence to the programmability offered by SDN, NFV, and NS can result in zero-touch network orchestration. An AI-enabled 6G network can benefit from robust techniques and capabilities in analyzing, learning, and optimizing concepts like management and orchestration \cite{3}. The presence of AI seems inevitable considering the complex and dynamic nature of 6G networks, which is due to the emerging new applications and fast-forwarding technology-oriented lifestyles. This has provided enough reason for the research society to suggest deploying various AI techniques in different sections of the network for multiple purposes. As a result, future wireless networks need to become “AI native.” In other words, in 6G, every relatively isolated network entity needs to be able to adjust to the momentary circumstances and work in the most proper manner. This initial intelligence results in a self-aware, self-adaptive, and self-interpretive network that emerges as agile and smart \cite{3,5}. By converting the conventional concept of the network to many independent yet connected intelligent network entities, 6G will transform 5G from the “Network as a Service” to the “AI as a service” concept \cite{0}.

While communication systems started as connection-oriented systems, the entire network is arranged to support such a communication model. However, intelligence and sensing are an indivisible part of 6G, and thus, the 6G system will be composed of distributed intelligent entities, each conducting an independent yet deterministic task. As a result, the communication mechanism in 6G needs to follow a task-oriented model instead of a monolithic connection-oriented system. A similar transition will happen in the network´s service-providing strategies. Even though we have already experienced tailored services for tenant requests, 6G will thoroughly change the network-centric services to the user-centric ones. While the term “user” represents a broad group, including robots, devices, machines, and groups of people in any size, a user-centric network model refers to a network mechanism that is user-configurable and user-controllable \cite{7}. Delivering the new characteristics of 6G starts with a transformative architecture designed with new technological enablers for an intelligent task-oriented and user-centric management, and it continues with a fully distributed and dynamic orchestration that manages automatic functional elements capable of providing on-demand services in various domains \cite{0}. 
Accordingly, we are motivated to consider a distributed and scalable system model that consists of multiple AI-enabled network domains with finite resources for NS and service provision in the next section.

\section{A multi-layer and decentralized AI-driven network system}
While the robustness and dynamic nature of 6G led us to consider decentralized and modular network system models, the management and orchestration of that model can become progressively complicated. Therefore, we are inspired to merge a decentralized and modular network system with a decentralized and automated management and orchestration system. As Fig.1 presents, we consider a system model consisting of two parts. 
%-----------------------------------------------
%\begin{figure}[t] 
%\centering
%\includegraphics[width=3.4in]{p4fig0.pdf}
%\vspace{-0.3cm}
%\caption{Network´s general framework}
%\vspace{-1.5em} 
%\label{figure0}
%\end{figure}
In the upper part, the management and orchestration section is responsible for the network´s overall management including slice orchestration, service provision, and resource allocation by mapping the requests to the service types and managing the domains. The second part is the AI-enabled infrastructure that includes network domains (edge domains, far-edge domains, etc.). It should be highlighted that even though the network is managed by the management and orchestration section, each network domain needs to be managed independently and through automation. This is because it is essential to avoid centralized management sections on which the entire network depends. Accordingly, we continue by studying the network domains closely since this work focuses on edge domain network slicing and automation.  
%-----------------------------------------------
\begin{figure*}[t] 
\centering
\includegraphics[width=6in]{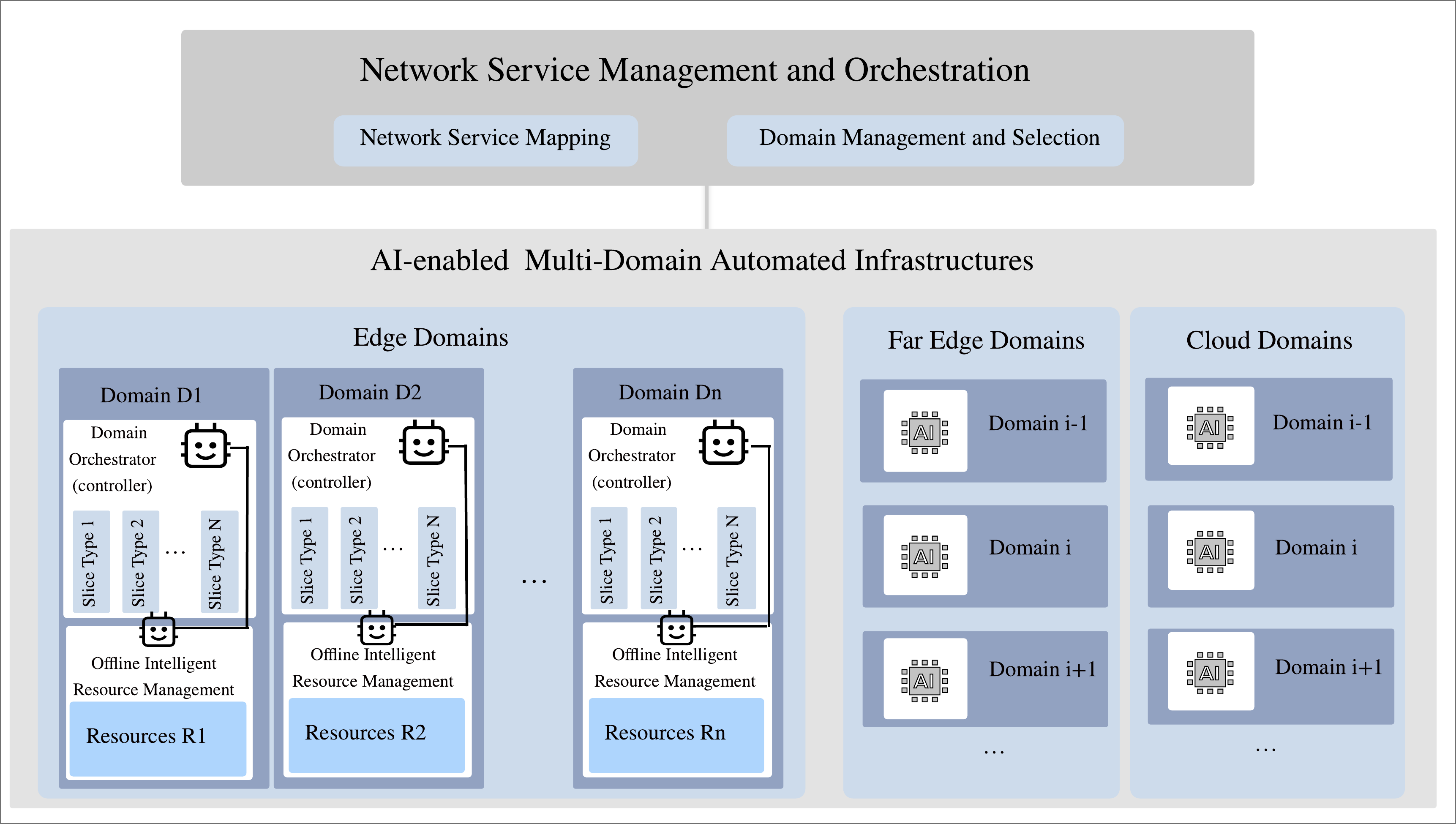}
\vspace{-0.3cm}
\caption{Network´s general framework with a close look to the edge domains}
\vspace{-1.5em} 
\label{figure1}
\end{figure*}
As Fig. 1 presents, each edge domain consists of resource pools that should be allocated to the users to provide a service. It is inevitable to have multiple numbers of each domain in the 6G networks. Therefore, multiple network edge domains that consist of resource pools can exist in the network. This emphasizes that each domain needs to be automated and manage the intra-domain resources independently in order for 6G networks to be scalable. As a result, we consider an intra-domain management and NS for each network edge domain in order to meet the robustness and scalability that are required for 6G networks. The domain orchestrator (controller) in Fig. 1 is responsible for assigning network slices for a required service type. It is clear that the domain controller can not be static in order to incorporate seamlessly in the overall dynamic management and orchestration and take the maximum benefit from the resource pools in the domains. Moreover, we target on the development of approaches that are efficient and scalable. Therefore, we will explore ML-based techniques that align with our problem in the next section and accordingly, we will propose an offline intelligent primary level of resource management in the domains.

%---------------------------

\section{Enabling technologies: RL, FL, TL and Unsupervised RL}
RL is a well-known ML-based technique for real-life problems that involve an intelligent agent capable of learning through interacting with an environment. The learning process occurs by taking actions and receiving rewards based on those actions. The RL-based technique that is applicable to decentralized management problems is MARL, which includes cooperative MARL, competitive MARL, or a combination of both. It is worth mentioning that the level of sharing information, coordination mechanisms, and observations can vary and thus result in other RL approaches. However, MARL usually suffers from scalability and complexity problems since multiple agents have a significant impact on the complexity level. 

FL, On the other hand, is a decentralized approach in which multiple servers in various locations collaborate to train an ML model without sharing their local data. In detail, each server or device trains a local model on its own data, and the central center only aggregates the model updates from the local participants. Although the efficient approach of only model-updates are transmitted to the central server is applied in FL, scalability concerns still exist. This arises from the requirement for a central node to aggregate updates, and as the number of participants increases, the complexity of managing them escalates. The heterogeneity of the devices and shared data is another challenge in addition to the communication overload. 

While distributed RL and FL are promising techniques, our focus is on minimizing complexity and maximizing efficiency in 6G network management and orchestration. In other words, complexity should be realistic for a system and management strategy to exist in real life. Given the inherent complexity of 6G networks and the anticipated challenges in NS and service provision, we are investigating management strategies to achieve optimal efficiency. Increasing efficiency and decreasing complexity happen in both the implementation and performance stages. Hence, ML-based methods that can adapt their trained model to new tasks and environments instead of requiring training for each and every task or locality are preferable for complex systems. 
Transfer Learning (TL) is an ML technique that leverages the knowledge that is gained from one task to improve the performance of a related task. This method consists of a pre-training phase that results in a trained model to be used for different tasks. Similarly, meta-learning is the concept of improving the learning process by making the models more adaptable to new tasks with minimal data and efficiency. 

Although TL appears efficient due to the multiple use of trained models for multiple tasks, we envision placing intelligent agents to automate the network domains. This is determined by the nature of the targeted problem. The domain controllers need to interact with the domain and take a sequence of actions in order to allocate available resources for providing network services. Considerably, we were inspired to benefit from a pre-training phase of unsupervised RL in order to have intelligent controllers while increasing the efficiency by taking advantage of previously learned information. We aim to add an offline level of unsupervised RL-based assignment skill discovery from which the domains' controllers can benefit. Adding a pre-training phase enables simple controllers to provide multiple service types in an efficient way.

\section{Our Methodology: Unsupervised RL-based Assignment Skill Learning}
Designing a reward system tailored to a specific task and environment can effectively guide an RL agent to learn a desirable behavior or policy. However, a hand-crafted reward is costly when trying to have a network of distributed intelligent entities with automation capabilities. Moreover, it is crucial to understand that although 6G is a heterogeneous network with a distributed and dynamic nature, the entire network follows some unified purposes. Considerably, unsupervised RL offers a pivotal enhancement. The advantage of applying unsupervised RL in 6G networks is the efficiency resulting from agents with unsupervised policies that can adapt to various tasks. However, the influence exerted by the environment on the agent remains significant. In order to have agents that can adapt to multiple tasks in multiple network domains in 6G, these agents need to be pre-trained in multiple reward-free environments that belong to the same domain but have different transition dynamics \cite{8}. 

Unsupervised learning in RL can be challenging for the agent since it has to collect useful data in a pre-training phase without any explicit reward. Recent research studies suggest that skill discovery is a promising pre-training approach for collecting diverse behaviors \cite{14,15,16}. Skill discovery means learning diverse behaviors with an intrinsic reward called skills \cite{15}. The discovered skills in the pre-training phase can later be fine-tuned or used in HRL. Skill discovery is considered a promising pre-training approach in order to learn diverse behaviors that are potentially useful for performing different tasks \cite{15}. This is due to the fact that skills are learned without any prior knowledge of the tasks \cite{16}. Accordingly, we were inspired to benefit from a phase of skill discovery to form the edge domain automation.

By defining a skill as a latent-conditioned policy that alters the state of the environment in a consistent way, the DIAYN algorithm \cite{16} seeks to identify a set of skills by maximizing the utility of this set rather than relying on a handcrafted reward function. A latent-conditioned policy is a policy that conditions its output, which is the selected action, on latent variables. In other words, while normal policies receive the states as input and give the actions as output, in DIAYN the policy receives a latent variable in addition to the states as input. Maximizing the utility of the skill set refers to maximizing the mutual information between states and skills. As a result, the empowerment of the controller agent whose action space is the discovered skills will be maximized. By maximizing the mutual information between states and skills, DIAYN incorporates skills controlling the visiting states. The information between the skill and the state is presented as the entropy of the skill after extracting the entropy of the skill given the state. Thus, DIAYN tries to maximize the mutual information in addition to the maximizing the entropy of the policy. 

DIAYN is implemented on top of the Soft Actor-Critic (SAC) algorithm  \cite{16}. SAC is a well-known algorithm for solving continuous tasks featuring continuous action space with precision and smoothness with complex system dynamics. As a result of having a stochastic policy, it assigns probabilities to each possible action and extracts samples from the probability. SAC is an entropy-regularized RL, which means that while it follows the goal of maximizing the sum of the rewards, it tries to keep the entropy of the policy as high as possible. Consequently, the stochastic policy will explore a wider range of actions. As SAC already maximizes the policy's entropy, DIAYN´s neural networks only need to maximize the predictability of a skill by knowing the state in order to find distinguishable skills \cite{16}.

As originally envisioned, in this framework, edge domain management starts with a pre-training phase of unsupervised interactions with the environment to discover useful skills \cite{8,15}. In the next phase, the primary policy can be used to either go through supervised training with a reward function (fine-tuning) \cite{8} or be employed as the lower level of HRL. 

\section{Generic unsupervised skill discovery for automation of distributed network domains: An energy-efficient approach}
In this work,  we propose an intelligent controller that is employed in each domain and benefits from an offline pre-training phase of unsupervised RL agent, as depicted in Fig. 2. 

\begin{figure*}[]
\centerline{\includegraphics[width=7in]{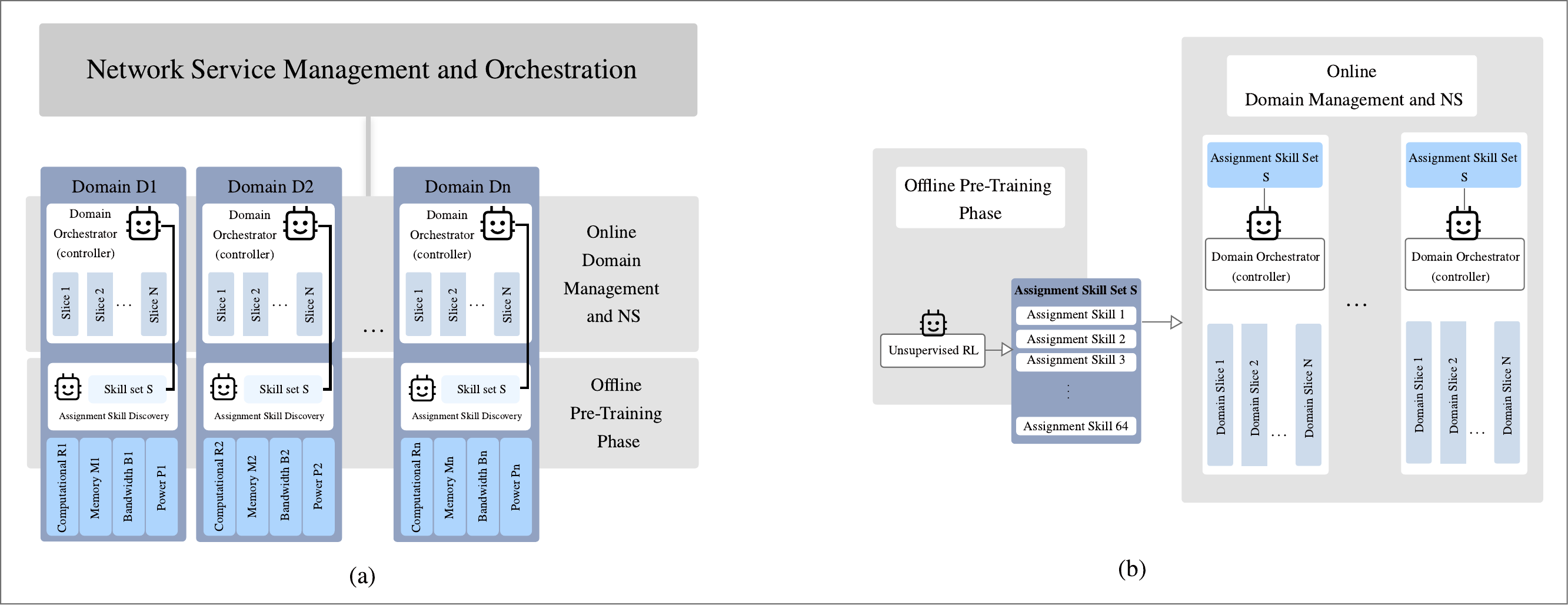}}
\vspace{-0.3cm}
\caption{Automation of edge domains: (a) Domains' management scheme in the system model, (b) A closer look to the offline and Online phases}
\vspace{-0.45cm}
\label{figure2}
\end{figure*}

Each domain considers four types of resource pools in a network domain including: computational resources (e.g., processors), power, bandwidth, and memory. On top of them, the unsupervised RL agent interacts with the resource pools directly. This level is called the pre-training level in which useful skills are discovered without supervision. The skills are later used for the domain controller. The controller´s tasks are different network services to be provided for a tenant. Thus, each skill is a set of assignments from four resource pools that will be used by the controller to provide a network slice for a tenant as a task. It is clear that the nature of the tasks is the same as a skill since a task is a set of resource assignments as well. Fig.2 (b) presents the domain management mechanism from left to right. In the offline phase the unsupervised RL agent discovers the useful assignment skills and since the domains are only different in details, the discovered skills can be used in all the domains. Adding a pre-training phase to discover the useful assignment tasks offers the following benefits:
\begin{itemize}
\item In each network slice, resources are allocated based on an upper limit, a lower limit, or a combination of both. Each resource pool can be considered as an interval of infinite numbers from which only specific numbers are useful for a network slice. Thus, the observation space of the controller is infinite, and the useful spaces are spars. On the other hand, skills discretize the infinite pool of resources, and while eliminating the need for complex controllers, they increase the probability of a successful slice assignment.
\item Since network domains are generally alike, discovering useful skills eliminates the need for a separate training process in every domain. Moreover, by defining the skills in percentage instead of the actual quantity of assignments, the discovered skills can adapt to domains with resource pools of different sizes.
\item In addition to having multiple slightly different domains, multiple service types exist in the network. This means that each controller faces more than one task, and thus, the traditional one-goal agent is not useful. However, by discovering useful skills, the nature of the problem changes. The controller only needs to select a sequence of skills from a finite set of skills until the constraints for the selected task are satisfied. Therefore, a simple and efficient controller can perform multiple tasks.
\item This approach proposes one offline pre-training phase that enables the domain to be automated by a simple controller, which offers enough efficiency for having multiple edge domains without any concern about complexity increment.
\item We envision the management of this decentralized system model to be in multi-levels. The discovered skills can be used in various ways, including by a simple controller, by fine-tuning the model, or even in HRL.
\end{itemize}

The following paragraphs details our implementation of the Unsupervised RL-Based Assignment Skill Discovery for Network Edge Domains' approach. 
In this scenario, the environment is the domain as shown in Fig.2. However, the only observable information for the offline pre-training phase are the resource pools. Each network slice encompasses communication (bandwidth), computational, memory, and power resources. At this level, our skills and states consist of four values representing the four mentioned resource types. In order to not limit our approach to one domain specifications, the action will choose values for each resource type in percentage. Consequently, the state and action space for the pre-training phase will be as follows:
\\
\\
State space= ${ [2-10]\% , [2-10]\%, [2-10]\%, [2-10]\%}$
\\
Action space = ${ [1-5]\% , [1-5]\%, [1-5] \%, [1-5] \%}$
\\

The observation space presents that only $10$ \% of each resource type can be used for a skill. In practice, no resource type can be zero for service provision. As a result, the lower bound of the state space is $2$ \%. The higher bound was set at $10$ \% to provide distinct states in every episode so that the agent is able to find distinguishable skills in different domains. This is because each episode starts with a different minimum and maximum percentage in the state space which can simulate the slight differences existing in different domains.

Since we consider the maximum percentage of each resource type for a skill to be $20$ \%, an upper bound higher than $10$ would limit the number of time steps in each episode. Action can be a continuous number between $1$ to $5$ \%. It is worth mentioning that the action space in this level consists of infinite real numbers, the discovered skills will be a finite number of discrete actions for the higher level controller. 

We set the agent to discover $64$ skills, a decision based on trials aimed at balancing comprehensive state-space coverage with discovering only the necessary and distinct skills. In this way, the infinite state space of the domain controller is reduced to a finite number of skills. Thus, while the controller state space will only include the most useful hand-picked assignment sequence, the action space will also be reduced to selecting a series of skills until the service is provided.

As mentioned in the previous Section, DIAYN uses an intrinsic reward system. Thus, we define the useful states as the final states so that when the unsupervised agent reaches a useful skill, the episode finishes. A final state is defined in two ways: when at least one resource type reaches the predefined maximum percentage ($20$ percent) or when the allocated resources represent a pattern. 
Considering the controller tasks, we need the discovered skills in the unsupervised offline phase to be the actions for tasks of providing different service types and intra-domain slices. Thus, if a skill matches any possible combination of an array consisting of $4$ binary numbers, the mentioned skill can be useful for the controller agent. Therefore, the states are normalized and compared to all combinations of a binary array with four elements. If the normalized state is close to any combination (with the relative tolerance of $1e-01$ and absolute tolerance of $1e-02$), that state is a final state. While the possible combinations form ¼ of the number of the skills ($2$\textsuperscript{$4$}), the rest of the final states have at least one resource in maximum use ($20$ percent). 

The agent was trained for 25,000 episodes with a $learning\hspace{0.1cm}rate$ of $1e-5$, a $gamma$ equal to $0.99$, $\alpha$ equal to $0.01$, and $\tau$ equal to $0.001$. 

Fig. 3 presents the $64$ skills by visualizing how much of each resource type the skills assign. In each skill from left to right, each chart bar shows the percentage of power, bandwidth, memory, and computational resources that the skill assigns. A close study of each skill confirms that all the skills are distinct and have an upper bound of $20$ percent. However, we need to confirm that replacing the state space of the controller with the set of discovered skills does not limit its access for an adequate service provision. Thus, we investigate the discovered skills in the entire space of the state space.
\begin{figure*}[t] 
\centering
\includegraphics[width=6 in]{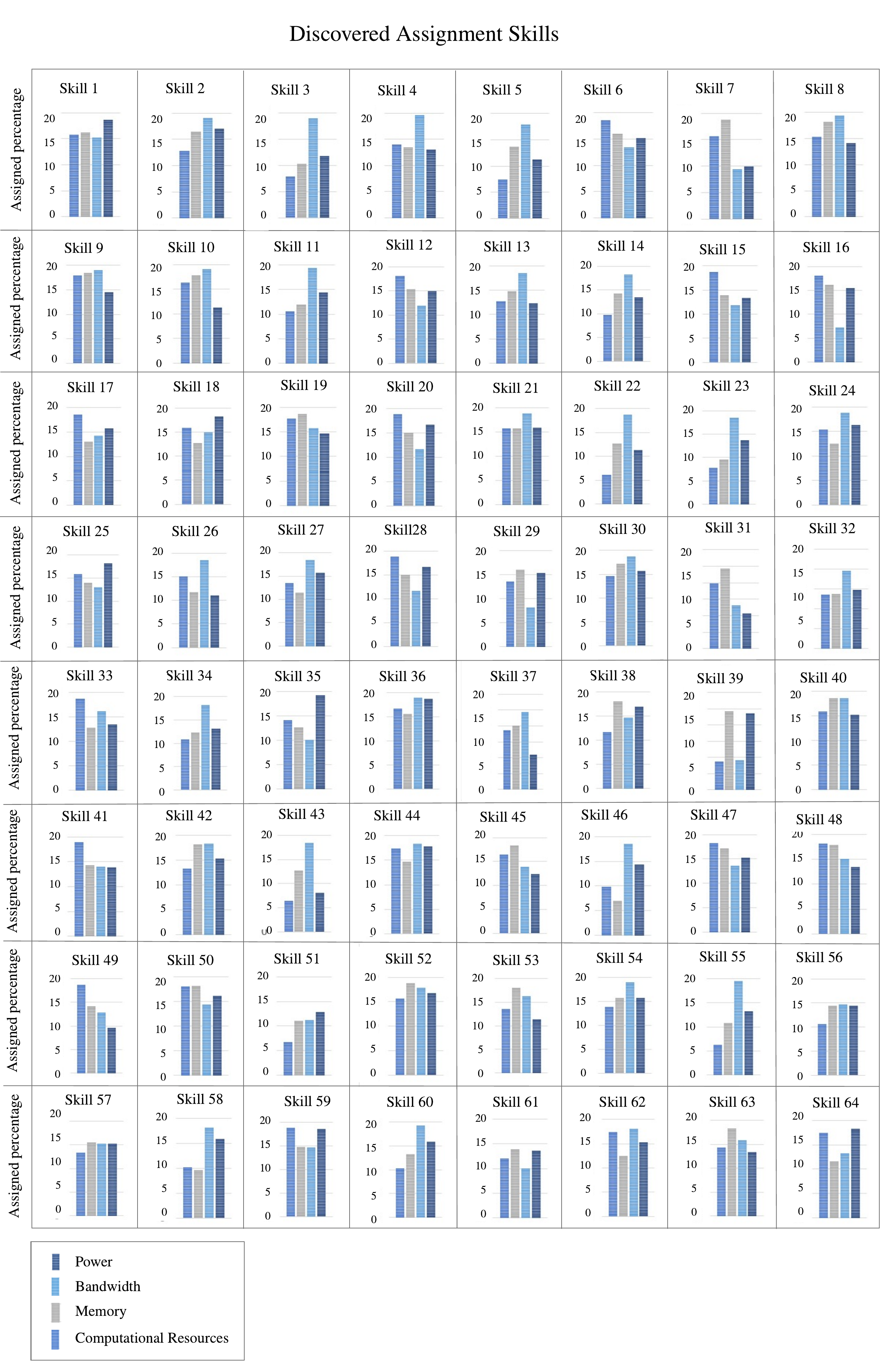}
%\vspace{-0.3cm}
\caption{Discovered Skills}
%\vspace{-1.5em} 
\label{figure3}
\end{figure*}
Fig. 4 demonstrates that a finite number of skills have extended across approximately $5$ to $20$ percent of the infinite state space of each resource pool. Thus, the unknown tasks of network slicing at the higher level can be managed by the minimum complexity, computational, and energy consumption in different domains since the finite number of skills cover the infinite space sufficiently. 
\begin{figure}[t] 
\centering
\includegraphics[width=2.5in, height=5in]{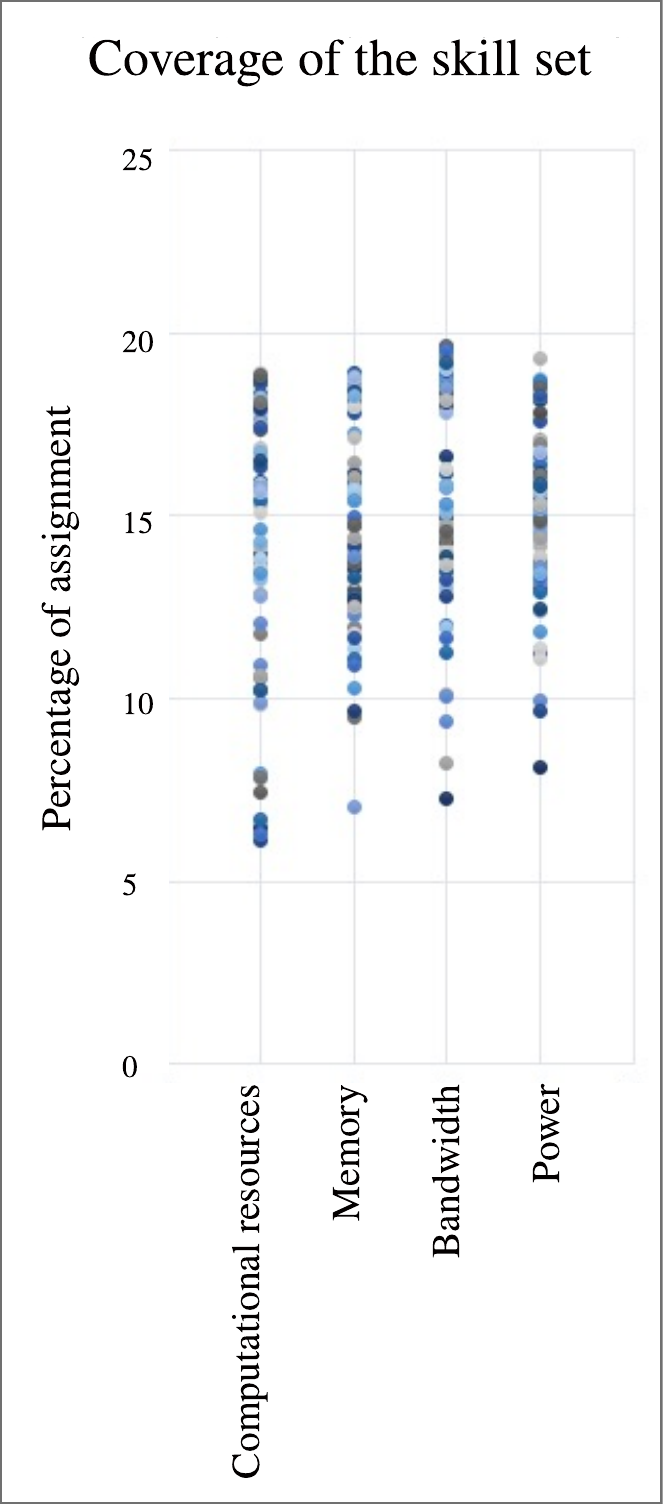}
\vspace{-0.3cm}
\caption{Coverage of the discovered skills across the resources of each domain}
\vspace{-1.5em} 
\label{figure4}
\end{figure}

\section{Prospective implementations and Advantages: HRL Orchestration}
Learning skills without supervision by intelligent agents have a revolutionary impact on practical applications. Since in the controller of network domains, the agent only receives rewards in some states, our network domains can be considered environments with sparse rewards. Learning useful skills in an offline pre-training level in environments with sparse goal states helps the agent in the online exploration. In addition, network service delivery can be considered a challenging task. This is the result of complex dependencies that exist in the sequence of assignments of various network resource types and blocks in order to deliver a service with predefined QoS. Thus, for these long-horizon tasks, if the skills perform as primitives for HRL, the global goal will be reached faster compared to training from scratch. Since traditional HRL uses skills that are learned by a known reward system, the set of skills and a meta-controller are trained jointly.  As a result, the controller affects the selected options by not allowing the exploration of options not good for that reward. However, as the hierarchical management scheme in our system needs intelligent agents that can manage the automation of slightly different network domains, the environment and the tasks are unfamiliar to the agent. Unsupervised skill discovery is task-agnostic and does not aim to solve a single known task \cite{16}. Thus, this approach meets the needs of our management scheme. Moreover, designing a reward function is a challenging task since the learning process completely depends on that.  Consequently, unsupervised skill discovery offers a more adaptable strategy and facilitates the technical work if it is used in an HRL orchestration.

\section{Energy Efficiency}
With the extensive growth of wireless network applications and their roles in our lives, green communication has become an important concern in telecommunication society. Green communication, as a broad term, refers to environment-friendly networking. In green communication, all the stages, from design and implementation to management, follow the goal of minimizing environmental impacts. While a wide range of design-related strategies, technologies, and considerations exist for improving the energy efficiency of wireless networks, energy efficient management and orchestration are also essential. 
Softwarization and virtualization provide us with the revolutionary opportunity of flexible and energy efficient resource management strategies that can reduce the idle time of hardware network items. Consequently, dynamic resource allocation and adoptable QoS offered by various service types prevent unnecessary resource and energy usage. Since AI will play a considerable role in software network management, it can contribute to energy efficiency-related concerns. However, programming the AI-enabled techniques require computational resources that utilize hardware resources. Developing scalable, generic, and efficient approaches are more efficient yet complicated and challenging in tailored service provision. In this work, we sought to balance dynamic and intelligent domain orchestration with energy efficiency. 
A close look at our automated modular orchestration confirms that a generic offline skill discovery primary phase minimizes the computational tasks that use hardware and energy. This is achieved by leveraging and adapting learned information across multiple domains. Moreover, generic skills simplify the duties of the higher level(s) of the orchestration and multiple service provision tasks can be executed in an uncomplicated process. Leveraging pre-trained models and results and reusing the knowledge gained from the domains reduces extensive training and energy consumption.

\section{Conclusions and future steps}
With 6G closer than ever, in this research work, we provided a background on what 6G will be. Some of the 5G characteristics were discussed since they are considered as enablers for 6G. Some fundamental transitions in the architecture and management from the previous generations were mentioned and discussed. After providing a short review of edge RL techniques, we presented a modular decentralized management system model. Then, we proposed an offline pre-training phase of unsupervised RL-based assignment skill discovery which results in increasing efficiency and scalability. 
While this work mentioned the application of unsupervised skill discovery in for the domain controller or in an HRL setting , the implementation was not presented. The offline unsupervised skill discovery can be utilized by a domain controller, by fine-tuning the pre-trained model, or in an HRL-based management in order to perform service-provision tasks. This work can be extended in future steps by implementing the higher level controllers.
 
\section*{Acknowledgement}
This work has received funding from the European Union under the ADROIT6G project (Grant agreement
ID: 101095363) and the H.F.R.I project ENABLE-6G (ID: 16294).

%  \section*{Biographies}
%\footnotesize
%\textbf{Navideh Ghafouri} is a PhD candidate at the Polytechnic University of Catalunya and working as a junior researcher in Iquadrat Informatica S.L. Her current focus is on the management of future wireless networks by deploying RL techniques.  \par
% \textbf{John S. Vardakas} received his Ph.D from the ECE Dept., University of Patras, Greece in 2012. His research interests include teletraffic engineering, performance analysis / simulation of communication networks and smart grids. \par
%\textbf{Kostas Ramantas} received his Ph.D in 2012 from CEID, University of Patras, Greece. His research interests are in modeling and simulation of network protocols; and scheduling algorithms for QoS provisioning. \par
%\textbf{Christos Verikoukis} received the Ph.D. degree from UPC, Barcelona, Spain, in 2000. He is currently an Associate Professor at the CEID, University of Partas, Greece.  \par

\end{document}